\documentclass[11pt,twoside]{article}
\usepackage{asp2010}

\begin{document}
\title{Total to central luminosity ratios of quiescent galaxies in MODS as an indicator of size evolution}
\author{Mohammad~Akhlaghi$^1$, Takashi~Ichikawa$^1$ and Masaru~Kajisawa$^2$
\affil{$^1$Astronomical Institute, Tohoku University, 6-3 Aramaki, Aoba-ku, Sendai 980-8578, Japan}
\affil{$^2$Center for Space and Cosmic Evolution, Ehime University, Bunkyo-cho 2-5, Matsuyama 790-8577, Japan}}

\begin{abstract}
Using the very deep Subaru images of the {\small GOODS-N} region, from the MOIRCS Deep Survey and images from the HST/ACS, we have measured the Luminosity Ratio ({\small LR}) of the outer to the central regions of massive ($M>10^{10.5}M_\odot$) galaxies at fixed radii in a single rest-frame for $z<3.5$ as a new approach to the problem of size evolution. We didn't observe any evolution in the median {\small LR}, Had a significant size growth occurred, the outer to central luminosity ratios would have demonstrated a corresponding increase with a decrease in redshift.
\end{abstract}


\section{Introduction}
The galaxy size is an inevitable part of any model of galaxy evolution, therefore observationally determining its evolution in the last $\sim12$Gyrs ($z\sim3.5$) will play a significant role in constraining the current models. As a review of some of the most recent studies; \cite{dad05} found that 4 of 7 massive passive galaxies at $z>1$ showed very small effective radii. \cite{dam09} claimed that there are no local counterparts to those compact objects. As a recent example \cite{cas11} found a 5 times number density increase in higher redshifts and a 0.5dex size increase. From a different perspective, \cite{vd09} found a high velocity dispersion compact galaxy at $z=2.186$. 

The results above were not left unchallenged tough and \cite{cen09} and \cite{cap09} failed to find any significant increase in the velocity dispersion. In a very interesting discovery, \cite{val10b} found a significant number of compact early type galaxies in the local universe and found similar number densities to those claimed for the highredshift compact galaxies, following those results, \cite{sar10,sar11} show that the number density of compact massive ETGs has stayed constant in the past 10Gyrs. In a recent study, \cite{ichikawa12}(in press) showed that a universal mass-size relation exists and that it is independent of redshift and slope.

The majority of these studies use the effective radii that is strongly volnerable to the object's outer regions, we believe that at least one of the reasons of such contrasting results can be in this volnerability, so in this paper and the subsequent (Akhlaghi et al. 2012, in preperation; here after A12) we will try to approach this problem from another point of view that will be further elucidated in \S3. In this paper the {\small WMAP} 7-year results \citep{wmap7} are used for the cosmological constants: $H_0=70$(km/s)/Mpc, $\Omega_{\Lambda}=0.73$ \& $\Omega_{M}=0.27$.

\section{Data}
In this analysis we have used the J, H \& $K_s$ images from the Multi Object Infra Red Camera and Spectrograph \citep[{\small MOIRCS},][]{suzuki08}, installed on the Casegrian focus of the Subaru telescope and B, V, r \& i images taken with {\small HST/ACS}. Detailed information regarding the reduction and analysis of the images and compilation of the catalog used here has been elaborated in \cite{kaj11}, hereafter {\small K11}. In this study we have only used the wide image and catalog results of {\small MODS} which show $5\sigma$ errors in one arcsecond diameter of 25mag/arcsec$^2$ in the $K_s$ band. It has been shown in {\small K11} that this catalog is $85\sim90$percent complete at 25 magnitues and so all the objects with $K_s$ band total magnitudes more than this value were removed from this analysis. Being the largest {\small PSF} value, all the images were {\small PSF}-matched to 0.6. {\small SED} fitting to obtain the observed mass and photometric redshift of the objects was preformed with the multiband photometry of the images explained above along with $3.6\mu{m}$, $4.5\mu{m}$ \& $5.8\mu{m}$ images from Spitzer/{\small IRAC}. Here we use the results of {\small GALAXEV} \citep{bs} with a Salpeter {\small IMF} \citep{salpeter}. The derived redshift showed excellent agreement with the spectroscopic redshift $\delta{z}/(1+z)=-0.011$.

\section{Analysis}
The new parameter used in this study and further elaborated in A12 is the Total to Central Luminosity Ratio ({\small LR}). The total luminosity is obtained from SExtractor's {\small MAG\_AUTO}, but the central radii is obtained from an aperture photometery with the size of 2.6kpc, which is the maximum physical size of our PSF during the cosmic history. Based on the definition of the effective radii, if {\small $LR=2$} then then $r_e=2.6kpc$ and independent of the galaxy morphology, if {\small $LR>2$}, then the effective radius is larger and if {\small $LR<2$} then the effective radii is smaller than that chosen value. The images were not deconvolved due to the simple reason that it adds to the noise we are trying so hard to avoid. 
\begin{figure}[tb]
\centering
\includegraphics[width=0.2\linewidth]{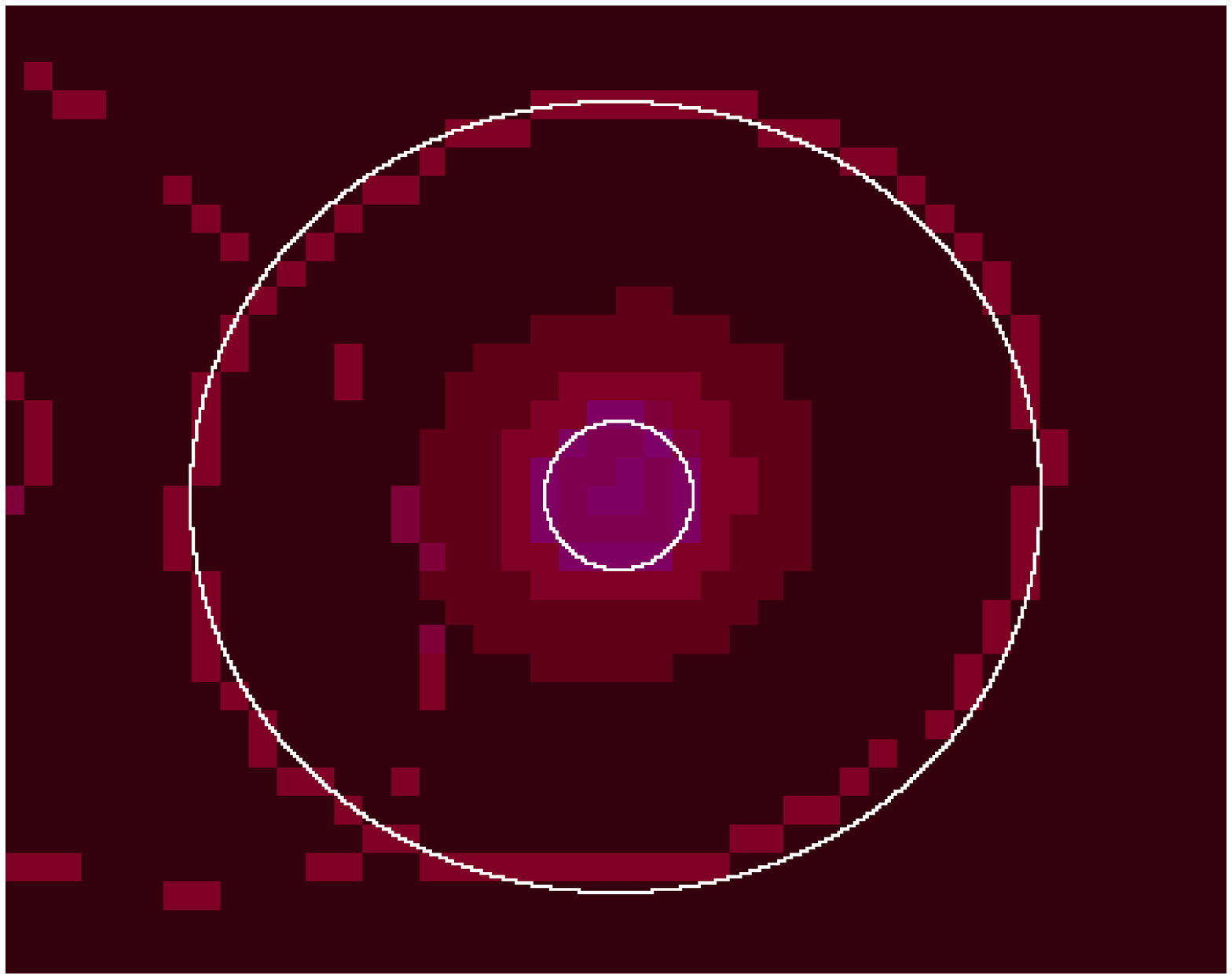}
\includegraphics[width=0.2\linewidth]{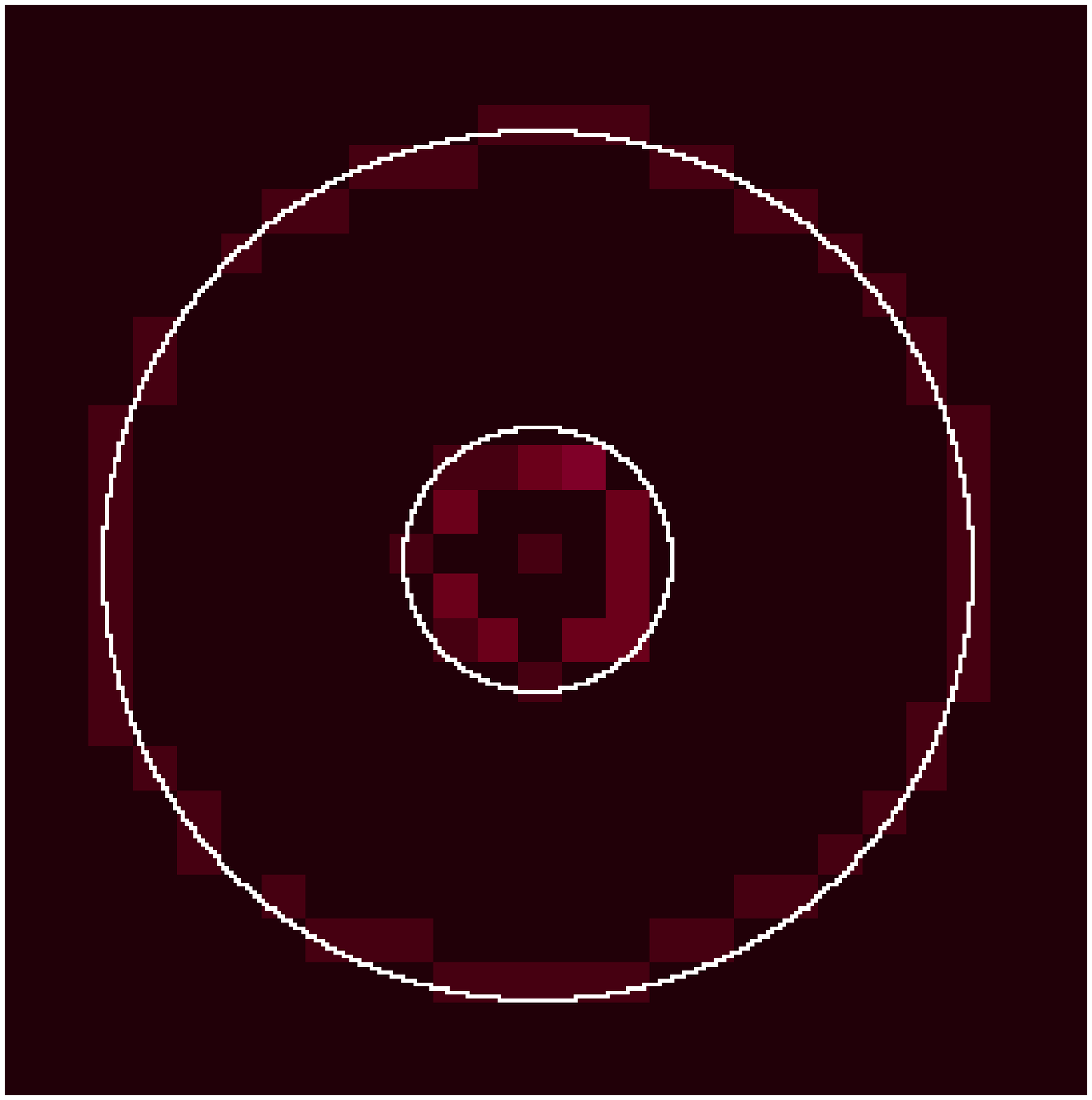}
\caption{Concept of Luminosity Ratio ({\scriptsize LR}) shown for two sample galaxies, the outer aperture is SExtractor's {\scriptsize MAG\_AUTO} and the inner one {\scriptsize MAG\_APER} set equal to 2.6kpc for the corresponding redshift. Left: A galaxy at $z\sim2.04$ with mass {\scriptsize $1.7\times10^{11}M_\odot$} with a V band 2.6kpc {\scriptsize LR} of 3.88. Right: A galaxy at $z\sim2.89$ with mass {\scriptsize $1.15\times10^{11}M_\odot$} with a V band 2.6kpc {\scriptsize LR} of 2.37.}
\label{sampleimages}
\end{figure}

In this short paper, we use the \cite{wil09} rest-frame color-selection technique to choose quiescent galaxies above {\small $M\approx10^{10.5}M_\odot$}. Since the stellar mass of the galaxy can best be represented by the restframe V images, we used the i, z, J, H \& K images for the different redshifts.

\section{Results \& Discussion}
\begin{figure}[t]
\centering
\includegraphics[trim = 18mm 50mm 17mm 5mm, clip, width=0.9\linewidth]{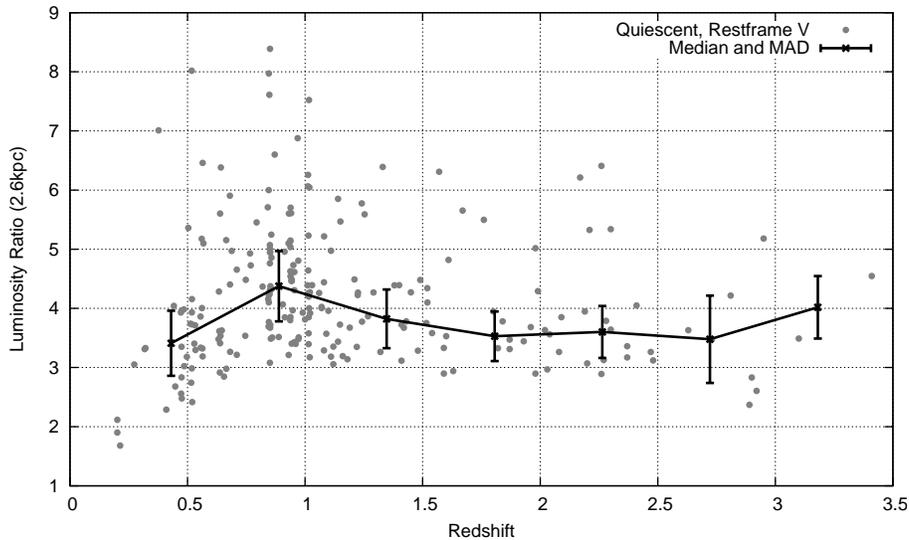}
\caption{Rest frame V band Luminosity ratio evolution for quiescent galaxies in {\small MODS} with physical aperture of 2.6kpc. {\small LR} for each galaxy is displayed as a gray dot, they are binned in 0.5 redshift bins, points connecting the lines are the median and the error bars are the Median Absolute Deviation ({\small MAD}).}
\label{QLRV}
\end{figure}

232 galaxies satisfied the quiescent and mass conditions we required and were detected by SExtractor in their desired image. {\small LR} was calculated for all galaxies with radii: 2.6kpc. The selected galaxies were binned to redshift bins of 0.5 and the median value of each bin was chosen as a representative; Fig. \ref{QLRV}. The dispersion is shown with the Median Absolute Deviation ({\small MAD}) as the error bars. 

The most salient result of Figs \ref{QLRV} is that it is basically a horizontal line and there is very slight variations of LR present. Put simply, Fig \ref{QLRV} shows that the median of the quiescent galaxies in {\small MODS} have not undergone any significant size evolution in the last 12Gyrs. Another interesting result that can be inferred from Fig \ref{QLRV} is that except a very low number of galaxies in the local universe, all values of {\small LR(2.6kpc)} are larger than 2, this means that all our galaxies have an effective radii larger than 2.6kpc. 

In some previous studies that have observed significant size evolution, the average value was chosen as a representative \citep[e.g.,][]{vd10}. As we see from Figs \ref{QLRV}, had we taken the average of the points, due the large dispersion and the outliers present in $z\sim1$, we would see a much more significant evolution in {\small LR} than we see now.

One important caveat of this current study is that the exact limits of our measurements and the exact relation it has with the effective radii have not been tested through simulated galxies embedded in our noise and {\small PSF}. Such simulations and analysis will be fully elaborated in A12.

\section{Summary}
As a conclusion it can be mentioned that based on a purely phenomenological approach to the problem of the size evolution of quiescent galaxies, and using a new parameter to study this evolution that will be completely scrutinized in A12, we find that quiescent galaxies in GOODS-North have not undergone any significant size evolution in the course of the last 12Gyrs.

\bibliography{akhlaghi.bbl}{}
\bibliographystyle{asp2010.bst}
\end{document}